\documentclass[aps,prl,superscriptaddress,reprint,showpacs,twocolumn,preprintnumbers,amsmath,amssymb,floatfix]{revtex4}

\usepackage{graphicx}
\usepackage{dcolumn}
\usepackage{bm}
\usepackage[utf8]{inputenc}

\begin{document}

\title{Magnetically-induced local lattice anomalies and low-frequency nematic fluctuations in the Mott insulator La$_2$O$_3$Fe$_2$Se$_2$}

\author{Riaz Hussain}
 \affiliation{Department of Physics, University of Pavia-CNISM, I-27100 Pavia, Italy}
 \author{Giacomo Prando}
 \email{giacomo.prando@unipv.it}
 \affiliation{Department of Physics, University of Pavia-CNISM, I-27100 Pavia, Italy}
\author{Sebastian Selter}
\affiliation{Leibniz-Institut f\"ur Festk\"orper- und Werkstoffforschung (IFW) Dresden, D-01171 Dresden, Germany}%
\affiliation{Institut für Festk\"orper- und Materialphysik, Technische Universit\"at Dresden, D-01069 Dresden, Germany}%
\author{Saicharan Aswartham}
\affiliation{Leibniz-Institut f\"ur Festk\"orper- und Werkstoffforschung (IFW) Dresden, D-01171 Dresden, Germany}%
\author{Bernd B\"uchner}
\affiliation{Leibniz-Institut f\"ur Festk\"orper- und Werkstoffforschung (IFW) Dresden, D-01171 Dresden, Germany}%
\affiliation{Institut für Festk\"orper- und Materialphysik, Technische Universit\"at Dresden, D-01069 Dresden, Germany}%
\affiliation{W\"urzburg-Dresden Cluster of Excellence ct.qmat, Technische Universit\"at Dresden, D-01062 Dresden, Germany}%
\author{Pietro Carretta}
 \email{pietro.carretta@unipv.it}
  \affiliation{Department of Physics, University of Pavia-CNISM, I-27100 Pavia, Italy}

\date{\today}

\begin{abstract}
We report a study of the Mott insulator La$_2$O$_3$Fe$_2$Se$_2$ by means of $^{139}$La nuclear quadrupole resonance (NQR). The NQR spectra evidence a single La site in the paramagnetic phase and two inequivalent La sites, La1 and La2, in the antiferromagnetically-ordered phase (N\'eel temperature $T_{\rm{N}} \simeq 90$ K). These two sites are characterized by different quadrupole couplings, indicative of distinct lattice and/or charge configurations likely segregated in domains. The dependence of the quadrupole coupling for La2 on temperature suggests that the magnetic order parameter drives the structural distortion. The nuclear spin-lattice and transverse relaxation rates, $1/T_1$ and $1/T_2$, evidence fluctuations in the paramagnetic phase with characteristic frequencies well below the Heisenberg exchange frequency, likely associated with nematic fluctuations.
\end{abstract}
\maketitle
The simultaneous occurrence of different microscopic interactions, such as the Coulombic correlations and the Hund coupling, together with the complex fermiology arising from several orbitals crossing the Fermi level make the electronic phase diagram of iron-based pnictides and chalcogenides extremely rich \cite{fivebands1,fivebands2,ReviewIBS}. The emergence of orbitally-selective Mott transitions, with a quasi-particle weight significantly depending on the involved Fe orbitals, and charge disproportionation are exemplary demonstrations of the intrinsic complexity of these materials \cite{OrbS,Dispr}. Iron-based pnictides and chalcogenides also host an electronic nematic order breaking the C$_4$ symmetry and leading to a tetragonal-to-orthorhombic lattice distortion \cite{Nematic,Structdist}. Although this phenomenology has been observed in several metallic compounds \cite{Bohmer1,Bohmer2}, evidences for nematic fluctuations and order in the less studied insulating iron-based materials are still missing.

La$_2$O$_3$Fe$_2$Se$_2$ is an insulator formed by alternating La$_2$O$_2$ and Fe$_2$OSe$_2$ stacked layers and displays antiferromagnetic order below the N\'eel temperature $T_{\rm{N}} \simeq 90$ K \cite{Mayer1992AngChem, Zhu2010prl, Free2010prb}. The electric crystalline field resolves the Fe orbital degeneracy and possibly drives the system close to an orbital-selective Mott transition, with a suggested relevant role of charge and spin fluctuations \cite{Giovannetti2015prb,DFTG}. Recently, La$_2$O$_3$Fe$_2$Se$_2$ has attracted further interest after the report of short-range orthorhombicity both in the paramagnetic and in the magnetic phase based on the analysis of neutron pair distribution function \cite{Neutron2}. The breaking of C$_4$ lattice symmetry has been associated with nematic fluctuations -- however, the local orthorhombicity is weakly dependent on temperature, at variance with what is expected for a magnetic driving mechanism.

In order to reach a deeper understanding of the local structure and of the low-frequency dynamics, we studied La$_2$O$_3$Fe$_2$Se$_2$ by means of $^{139}$La nuclear quadrupole resonance (NQR). NQR is extremely sensitive to the changes in the local charge distribution induced by a modification of symmetry \cite{Imai, Carretta2020rnc}, and does not require the application of an external magnetic field as in the case of nuclear magnetic resonance (NMR). This is particularly useful when the experiments are performed on powder samples where the broad NMR lines require a more demanding analysis of the results \cite{Gunther2014prb}. In this letter, we report clear evidence of two distinct La sites, La1 and La2, only in the magnetic phase. La1 and La2 are characterized by different quadrupolar couplings, namely, by a non-equivalent local structure and/or charge distribution, suggesting the stabilization of the nematic phase in segregated domains for $T < T_{\rm{N}}$. The temperature dependence of the quadrupole coupling for the La2 site indicates that the magnetic order parameter is the origin of the structural distortion. Additionally, measurements of the $^{139}$La NQR transverse relaxation rate $1/T_2$ confirm that low-frequency dynamics develop just above $T_{\rm{N}}$, with characteristic frequencies well below the Heisenberg exchange frequency, consistently with the onset of nematic fluctuations in the paramagnetic state which progressively slow down as the temperature is decreased. The characteristic correlation time describing these fluctuations shows an activated behaviour, similarly to what was observed in the normal phase of iron-based superconductors and in prototypes of the $J_1-J_2$ model on a square lattice ($J_1$ and $J_2$ quantifying the exchange interactions among nearest neighbour and next-nearest neighbour magnetic moments, respectively) \cite{Bossoni2016prb, Bossoni2013prb, Li2VOsiO4}.
\begin{figure}[t!] 
    \centering
    \includegraphics[width=0.45\textwidth]{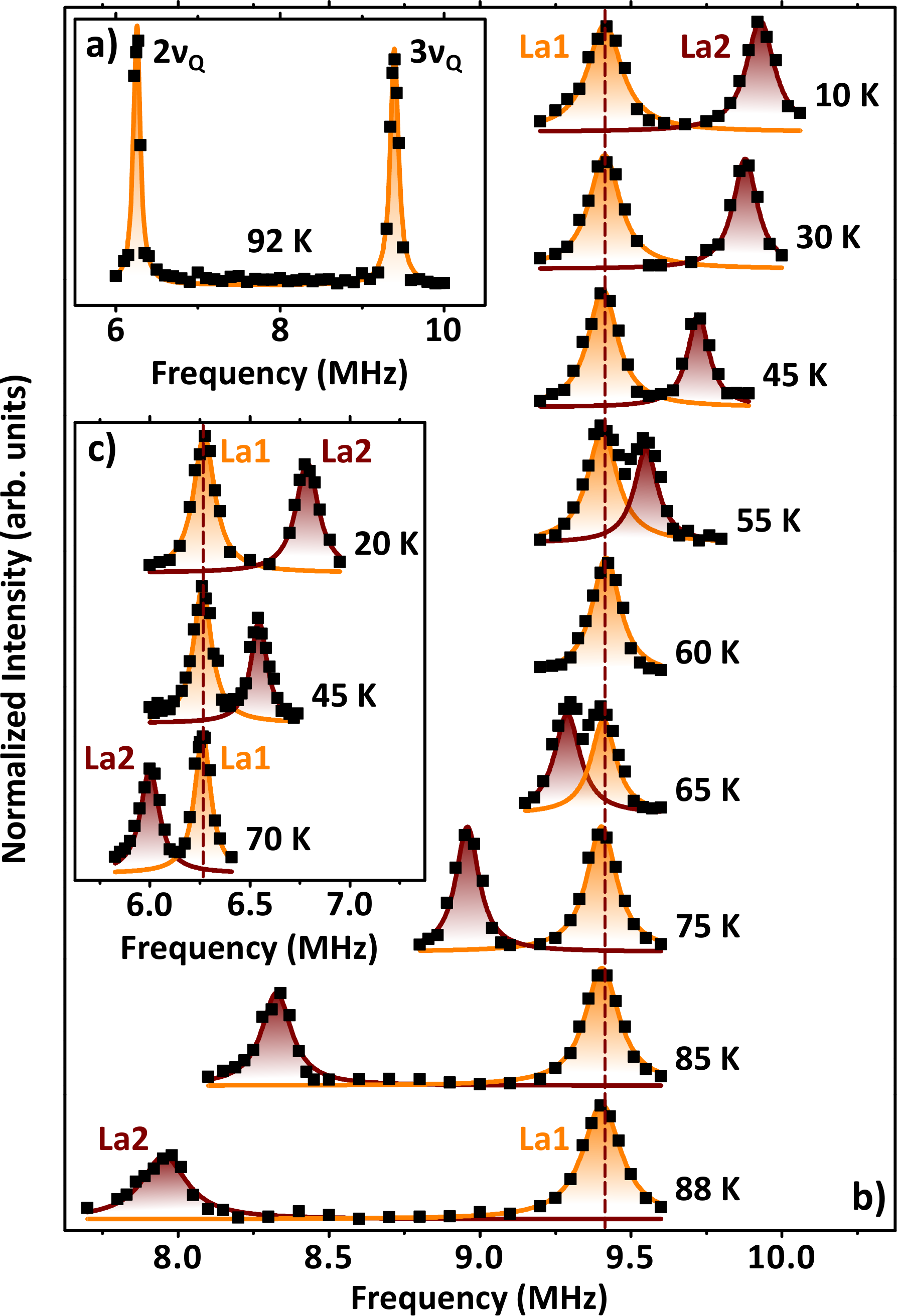}
    \caption{\label{spectra} Panel a) shows the NQR spectrum at $92$ K in the frequency range between $2\nu_{_{\rm{Q}}}$ and $3\nu_{_{\rm{Q}}}$, proving the disappearance of the La2 signal for $T \geq T_{\rm{N}}$. NQR spectra at several temperatures below $T_{\rm{N}}$ are shown in panels b) and c) ($3\nu_{_{\rm{Q}}}$ and $2\nu_{_{\rm{Q}}}$, respectively). Orange and dark-red lines are best-fitting Lorentzian functions for La1 and La2, respectively.}
\end{figure}

We performed the $^{139}$La NQR measurements on a La$_2$O$_3$Fe$_2$Se$_2$ polycrystalline sample using a broadband Tecmag Apollo spectrometer \footnote{See the supplemental material, which includes Refs.~\cite{Mayer1992AngChem, Recovery2, Gunther2014prb, Neutron2}, for detailed information on the sample synthesis and characterization as well as on the $^{139}$La NQR spectra and relaxation measurements.}. We measured the spectra by recording the amplitude of the echo signal after a Hahn echo pulse sequence $\left(\frac{\pi}{2} - \tau_{\rm{e}} - \pi\right)$ as a function of the irradiation frequency. We derived the spin-lattice relaxation time $T_1$ by recording the recovery of nuclear magnetization $M(\tau)$ after a saturation pulse sequence $\left(\frac{\pi}{2} - \tau - \frac{\pi}{2} - \tau_{\rm{e}} - \pi\right)$ and fitting the data with the functions appropriate for a nuclear spin $I = \frac{7}{2}$ in NQR \cite{Recovery1, Recovery2}. We quantified the transverse relaxation time $T_2$ from the decay of the echo amplitude after a Hahn echo sequence as a function of the inter-pulse delay, as well as with a Carr-Purcell-Meiboom-Gill (CPMG) sequence, which is particularly useful to evidence low-frequency dynamics \cite{CPSlichter}. In the CPMG pulse sequence, after the initial $\frac{\pi}{2} - \tau_{_{\rm{CPMG}}} - \pi$ pulse protocol, a series of refocusing $\pi$ pulses are applied at $2 \tau_{_{\rm{CPMG}}}$ intervals.
\begin{figure}[t!] 
	\centering
	\includegraphics[width=0.45\textwidth]{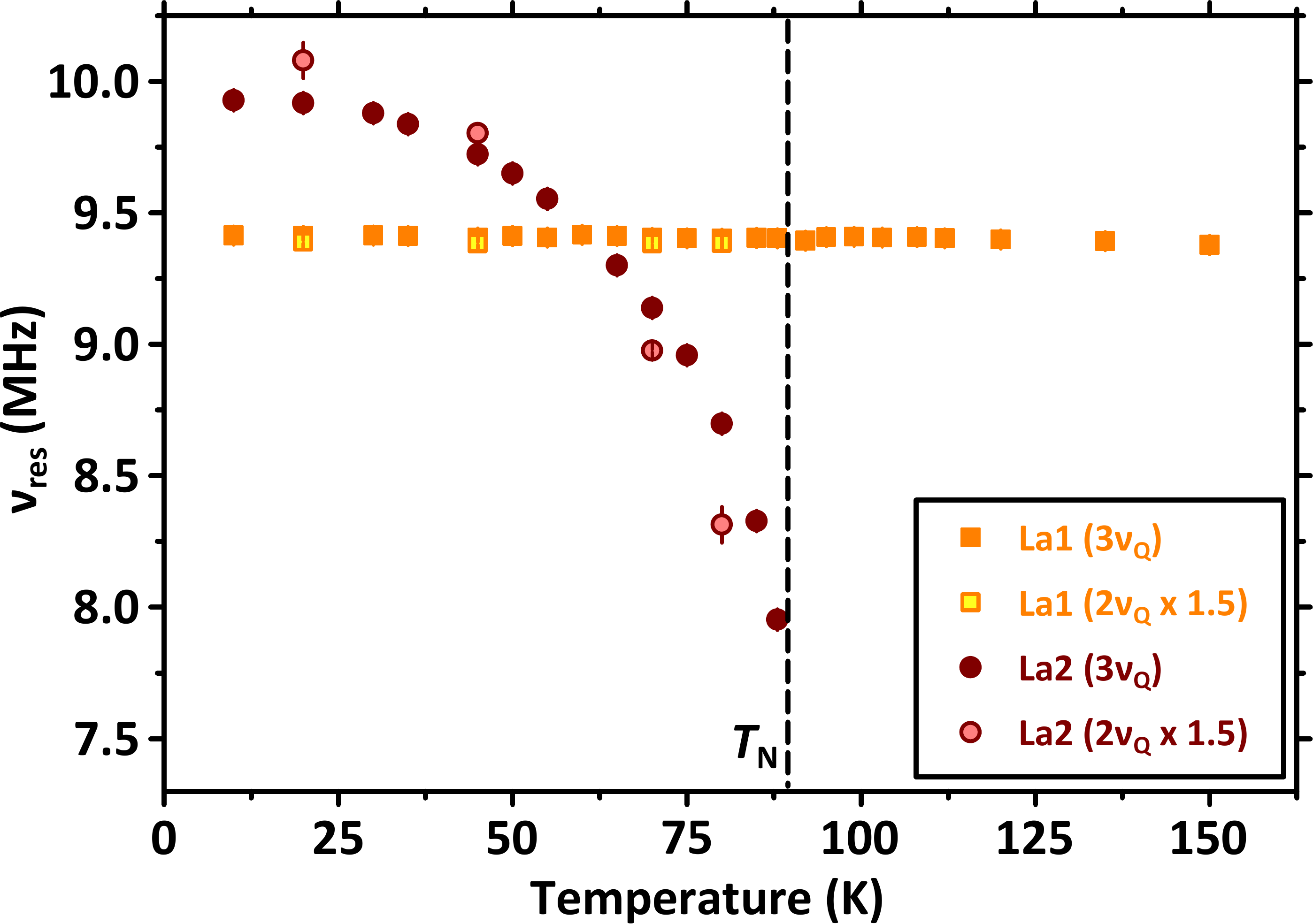}
	\caption{\label{nuT} Temperature dependence of the central resonance frequency of La1 and La2 signals (filled squares and circles, respectively) at $3\nu_{_{\rm{Q}}}$. Open squares and circles represent the central resonance frequency of La1 and La2 signals (respectively) at $2\nu_{_{\rm{Q}}}$ multiplied by a factor $1.5$. The dashed vertical line locates $T_{\rm{N}}$.}
\end{figure}

The resonance frequency in NQR spectra is determined by the interaction of the nuclear electric quadrupole moment $eQ$ with the local electric field gradient (EFG). Given the symmetry of the surrounding charge distribution in La$_2$O$_3$Fe$_2$Se$_2$, in the paramagnetic phase the NQR spectrum is characterized by three lines associated with the $m_I= \pm \frac{1}{2} \leftrightarrow \pm \frac{3}{2}$, $\pm \frac{3}{2} \leftrightarrow \pm \frac{5}{2}$ and the $\pm \frac{5}{2} \leftrightarrow \pm \frac{7}{2}$ transitions \cite{Abragam}. Here, $m_I$ is the component of the nuclear spin along the EFG main axis $Z$. These transitions are centred at $\nu_{_{\rm{Q}}}$, $2\nu_{_{\rm{Q}}}$ and $3\nu_{_{\rm{Q}}}$ frequencies, respectively, where $\nu_{_{\rm{Q}}}= eQV_{ZZ}/14h$, with $V_{ZZ}$ the main EFG component. Although we performed most of the measurements by irradiating the 3$\nu_{_{\rm{Q}}}$ line, the measurements at the 2$\nu_{_{\rm{Q}}}$ transition have proven very useful for the interpretation of our results.

For $T \gg T_{\rm{N}}$, the 3$\nu_{_{\rm{Q}}}$ line is centred at $\sim 9.4$ MHz (see Fig.~\ref{spectra}a), in excellent agreement with previous estimates of $\nu_{_{\rm{Q}}}$ based on NMR measurements \cite{Gunther2014prb}. The central resonance frequency is temperature independent within a few per mil down to $10$ K. Hereafter, we shall refer to this NQR signal as La1. Remarkably, a second NQR peak (La2) arises for $T \ll T_{\rm{N}}$ (see Fig.~\ref{spectra}b). Although the temperature dependence of the central resonance frequency of the La2 signal is akin to that of the magnetic order parameter, its frequency shift cannot be associated with the internal hyperfine field generated by Fe magnetic moments. First of all, the temperature dependence of the La2 peak frequency is less pronounced than that of the order parameter (see Fig.~6 in the supplemental material). Moreover, the internal field should lead to a splitting of the $^{139}$La NQR peaks \cite{Borsa} and the fact that neither La1 nor La2 peaks split below $T_{\rm{N}}$ rather indicates that the internal hyperfine field at $^{139}$La is weak (below $\sim 100$ G) and/or parallel to the $ab$ plane \cite{Gunther2014prb}. We stress that the intensity of La1 and La2 peaks is similar and that other nuclei, including $^{77}$Se, should have an intensity of the zero-field NMR lines more than an order of magnitude smaller \footnote{According to commercial NMR tables the absolute sensitivity at constant field, referred to $^1$H, is $5.91\times 10^{-2}$ for $^{139}$La and $5.25\times 10^{-4}$ for $^{77}$Se. Working at constant frequency the difference is even larger. On the other hand, the fact that one is irradiating just one of the three {NQR} transitions of $^{139}$La causes a reduction of the intensity of $3\nu_{_{\rm{Q}}}$ transition by a factor of the order of the unity.}.
\begin{figure}[t!] 
    \centering
    \includegraphics[width=0.45\textwidth]{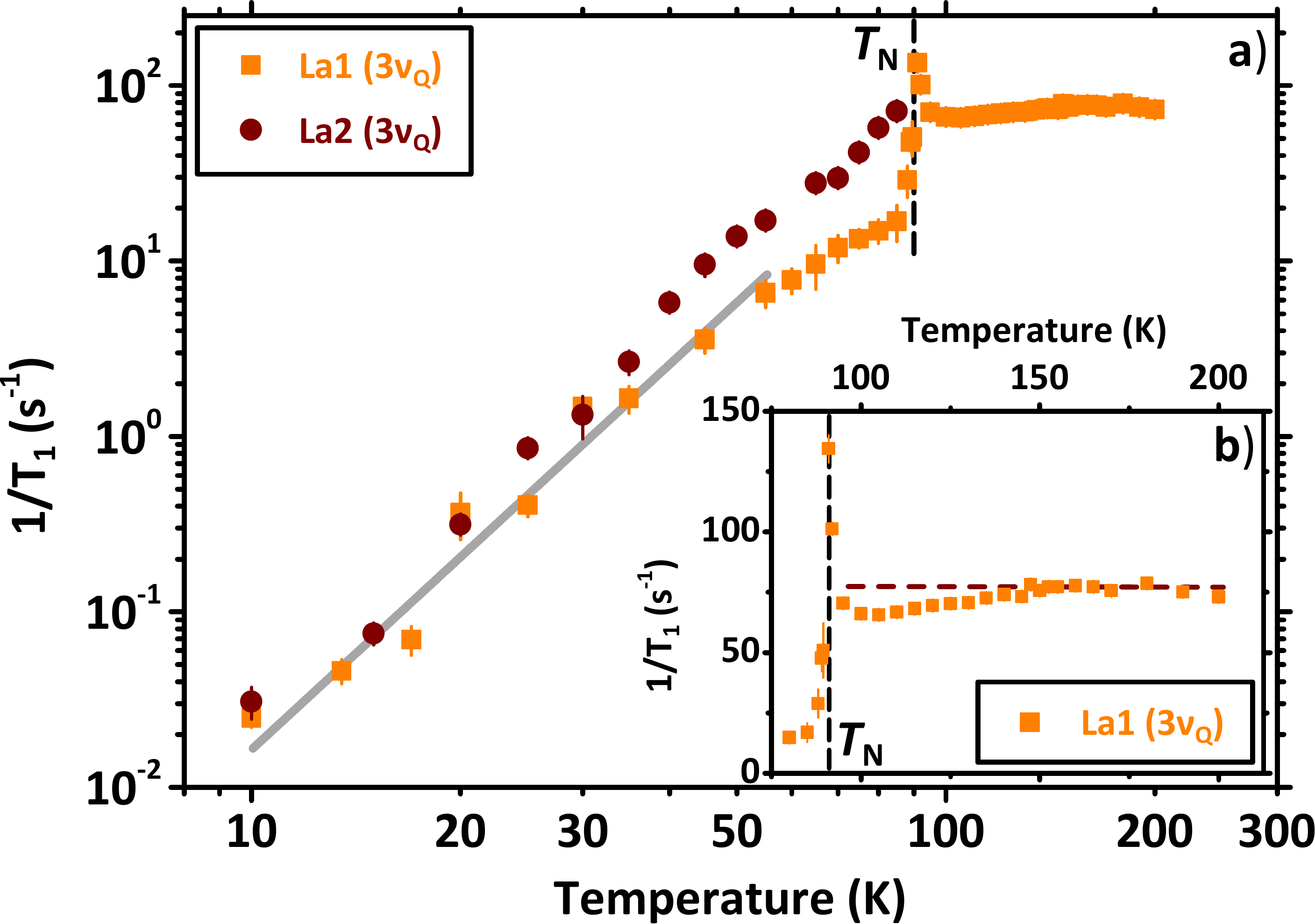}
    \caption{\label{invT1} Panel a) shows the spin-lattice relaxation rate for La1 and La2 (filled squares and circles, respectively) at $3\nu_{_{\rm{Q}}}$ as a function of temperature (log-log scale). The gray line highlights the $1/T_1 \propto T^{3.4}$ power law behaviour discussed in the text. Panel b) highlights the temperature dependence of $1/T_1$ for the La1 signal in the paramagnetic regime (lin-lin scale). The dashed horizontal line is a guide to the eye. In both panels, the dashed vertical line locates $T_{\rm{N}}$.}
\end{figure}

Remarkably, the behaviour of the $2\nu_{_{\rm{Q}}}$ transition is similar to what observed at $3\nu_{_{\rm{Q}}}$ (see Fig.~\ref{spectra}c). By scaling the frequency of the La1 and La2 peaks at 2$\nu_{_{\rm{Q}}}$ by a factor $1.5$ we find a good matching with the corresponding $3\nu_{_{\rm{Q}}}$ frequencies (see Fig.~\ref{nuT}), confirming that the signals are associated with $^{139}$La. These observations also confirm that the frequency shift of the La2 peak is determined by the quadrupolar interaction, i.e., its temperature dependence is determined by the evolution of the local structure and/or charge distribution around $^{139}$La nuclei.

Our results indicate that one of the La sites remains basically unaffected throughout the whole explored temperature range (La1) while the other site (La2) likely arises from a structural distortion driven by the onset of short-range nematic order. The presence of two distinct La sites was already pointed out in the NMR studies by G\"unther \textit{et al.} \cite{Gunther2014prb}. However, they concluded that two magnetically inequivalent La sites were present with different local hyperfine fields, whereas here we show that the difference is in the local structure or charge distribution around La1 and La2 sites. The M\"ossbauer measurements carried out by G\"unther \textit{et al.} also evidence a small change in the EFG at Fe site developing below 150 K \cite{Gunther2014prb}, hence it is possible that a change in the local structure is already taking place well above $T_{\rm{N}}$ and that Fe nuclei in M\"ossbauer experiments probe an average of the local configurations probed by La1 and La2 sites. This anomalous behaviour seems more likely associated with the presence of distorted and undistorted domains or a charge-density phase separation as predicted for iron-based materials by de' Medici and co-workers \cite{deMedici2017prl, CDphaseS}. Remarkably, the dependence of the central resonance frequency for the La2 signal is akin to that of the order parameter \cite{Neutron2}, suggesting that the magnetic order parameter is the driving force for the distortion. The presence of domains could also explain why neutron pair distribution function indicates a significant short range orthorhombicity and an overall long-range tetragonal structure \cite{Neutron2}.

We now turn to the discussion of the low-frequency dynamics associated with the onset of nematic fluctuations at $T \gg T_{\rm{N}}$. In the paramagnetic phase, we found that the recovery of the $^{139}$La nuclear magnetization by irradiating at 2$\nu_Q$ is two times faster than the recovery by irradiating at 3$\nu_Q$, consistently with a relaxation mechanism driven by spin fluctuations and not by EFG fluctuations \cite{Recovery2}. In the case of relaxation driven by electric quadrupole interactions, the recovery at the two frequencies should be comparable~\cite{Campana2000epj}. The temperature dependence of the spin-lattice relaxation rate for the La1 site at $3\nu_{_{\rm{Q}}}$ is shown on a log-log plot in Fig.~\ref{invT1}a. $1/T_1$ is temperature-independent for $T \gg T_{\rm{N}}$ (see Fig.~\ref{invT1}b), as expected for an insulating paramagnet \cite{Moriya}. Upon decreasing temperature, it first shows a weak decrease below about 150 K and then a sharp increase very close to $T_{\rm{N}}$. Below $T_{\rm{N}}$, $1/T_1$ shows a significant decrease by orders of magnitude with a low- temperature $1/T_1\propto T^{3.4}$ power law, close to the $T^3$ law typical of three dimensional antiferromagnets for $T_{\rm{N}} \gg T \gg \Delta$ \cite{Pincus}, where $\Delta$ is the spin-wave gap. For $T \ll T_{\rm{N}}$, the magnitude and behaviour of the spin-lattice relaxation rate is similar for both La1 and La2 sites, as shown in Fig.~\ref{invT1}a. However, the $1/T_1$ for the La2 site is enhanced with respect to that of the La1 site just below $T_{\rm{N}}$, suggesting a possible contribution from nematic fluctuations.

The observed behaviour for La1 is analogous to the one derived from NMR measurements, however we outline important differences. In NMR a very slight increase in $1/T_1$ was reported upon decreasing temperature from 150 K to $T_{\rm{N}}$ \citep{Neutron2}, while here we observe a decrease in $1/T_1$ \footnote{Freelon et al. \citep{Neutron2} report $1/T_1T$ data where the $1/T$ factor yields an additional enhancement of the data at lower temperature. We further remark that since La$_2$O$_3$Fe$_2$Se$_2$ is an insulator there is no reason to report $1/T_1T$ as commonly done in metals where this quantity is used to evidence the departure from a Fermi liquid behaviour.}. A possible explanation is that in the NMR experiments carried out by G\"unther \textit{et al.} \cite{Gunther2014prb}, given the chosen irradiation frequency, they were probing local field fluctuations in a plane tilted by about 36 degrees from the c-axis \cite{Abragam}. However, with NQR, we probe fluctuations perpendicular to the $Z$ axis which is parallel to the crystallographic $c$ axis -- in other terms, we probe fluctuations in the $ab$ plane. Hence, the observed behaviour could indicate a progressive decrease of in-plane fluctuations as the temperature decreases. At the same time, for a relaxation mechanism driven by spin fluctuations:
\begin{equation}
\frac{1}{T_1}= \frac{\gamma^2}{2}\sum_{\vec{q},\alpha=x,y,z} |A_{\vec{q}}|^2 S_{\alpha\alpha}(\vec{q}, \omega_R)
\end{equation}
where $\gamma$ is the nuclear gyromagnetic ratio, $S_{\alpha\alpha}(\vec{q}, \omega_R)$ are the components of the dynamical structure factor at the resonance frequency $\omega_R$ yielding a fluctuating hyperfine field in the $ab$ plane and $|A_{\vec{q}}|^2$ is the form factor describing the hyperfine coupling between $^{139}$La nuclei and the spin excitations at wavevector $\vec{q}$. Since for La1 the form factor filters out antiferromagnetic fluctuations, an initial decrease in $1/T_1$ is expected as the spin correlation length starts to increase, similarly to what is found in other two-dimensional antiferromagnets \cite{CFTD, SrZnVOPO}.
\begin{figure}[t!] 
    \centering
    \includegraphics[width=0.45\textwidth]{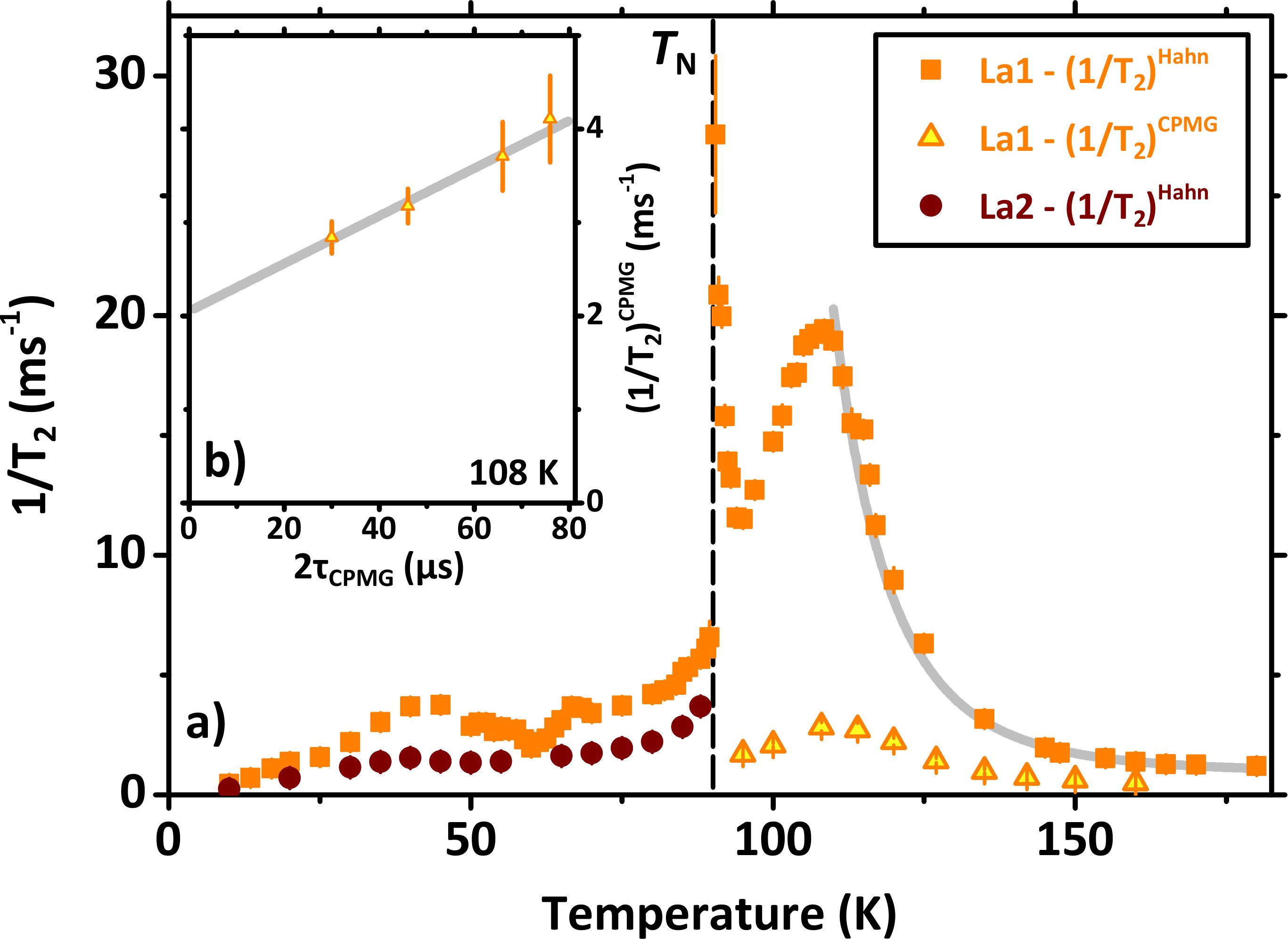}
    \caption{\label{invT2} Panel a) shows $\left(1/T_2\right)^{\rm{Hahn}}$ for La1 and La2 (filled squares and circles, respectively) as a function of temperature. The gray line is a best-fit according to an activated Arrhenius law. The open triangles show the temperature dependence of $\left(1/T_2\right)^{\rm{CPMG}}$ at fixed pulse separation ($2\tau_{_{\rm{CPMG}}} = 24$ $\mu$s). Panel b) shows the effect of increasing the pulse separation on $\left(1/T_2\right)^{\rm{CPMG}}$. The gray line is a linear best-fit to data.}
\end{figure}

Although the temperature dependence of $1/T_1$ for $T \gg T_{\rm{N}}$ is mainly driven by the correlated spin dynamics at frequencies of the order of Heisenberg exchange frequency, the temperature dependence of $1/T_2$ is determined by low-frequency (MHz range) dynamics likely associated with nematic fluctuations. The temperature dependence of $\left(1/T_2\right)^{\rm{Hahn}}$, measured with a Hahn echo sequence at the $3\nu_{_{\rm{Q}}}$ transition for the La1 site is shown in Fig.~\ref{invT2}a. Upon decreasing temperature, $\left(1/T_2\right)^{\rm{Hahn}}$ starts increasing around $150$ K and shows a clear maximum at $108$ K, well above $T_{\rm{N}}$, and shows a divergence at $T_{\rm{N}}$ associated with the critical slowing down of spin fluctuations. Below $T_{\rm{N}}$, $\left(1/T_2\right)^{\rm{Hahn}}$ for La1 and La2 sites slowly decreases with decreasing temperature, except for a small bump at the La1 site around 40 K whose origin should be investigated further.

In order to understand if the peak in $1/T_2$ at $108$ K marks a phase transition or a progressive slowing down of the fluctuations to very low-frequencies, we carried out a study of the echo decay with the CPMG sequence. The CPMG sequence makes it possible to probe the time evolution of the fluctuations at La1 site over a shorter timescale $\tau_{_{\rm{CPMG}}}$. If $\tau_{_{\rm{CPMG}}} \ll \tau_c$, where $\tau_c$ is the characteristic correlation time describing the dynamics, the fluctuations do not have enough time to affect the echo decay and $\left(1/T_2\right)^{\rm{CPMG}}$ decreases. This is indeed the behavior we observe which reveals that the fluctuations slow down with a $\tau_c$ of the order of tens of $\mu$s around $110$ K (see Fig.~\ref{invT2}b). Above the peak temperature, where the fluctuations are faster, one expects that $1/T_2 \simeq \gamma^2 \langle \Delta h_Z^2 \rangle \tau_c$, with $\langle \Delta h_Z^2 \rangle$ the mean square amplitude of the field fluctuations along the $Z \parallel c$ axis. We fit the temperature dependence of $1/T_2$ for $T > 110$ K with an Arrhenius law (see Fig.~\ref{invT2}a), indicating that $\tau_c \propto \exp(E_A/T)$, with $E_A= 990\pm 90$ K.

The behavior of $1/T_2$ is very similar to what observed in the normal phase of Ba(Fe$_{1-\rm{x}}$Rh$_{\rm{x}}$)$_2$As$_2$ \cite{Bossoni2016prb, Bossoni2013prb} and in Li$_2$VOSiO$_4$ \cite{Li2VOsiO4}, a prototype of the $J_1- J_2$ model on a square lattice. In both cases, dynamics at frequencies orders of magnitude lower than the Heisenberg exchange frequency contribute to $1/T_2$. In particular, a clear difference was reported in the $1/T_2$ measured with a Hahn echo and with a CPMG sequence in Rh-doped BaFe$_2$As$_2$ \cite{Bossoni2016prb}. Moreover, the low-frequency dynamics are characterized by an activated correlation time $\tau_c\propto \exp(E_A/T)$ both in Ba(Fe$_{1-\rm{x}}$Rh$_{\rm{x}}$)$_2$As$_2$ and in Li$_2$VOSiO$_4$, with $E_A$ of the order of hundreds of Kelvin degrees for Ba(Fe$_{1-\rm{x}}$Rh$_{\rm{x}}$)$_2$As$_2$ in the presence of a magnetically ordered ground-state. In a simple $J_1-J_2$ model the energy barrier characterizing the nematic fluctuations is the one separating degenerate collinear spin ground-states, with $E_A$ depending on $J_1$, $J_2$ and on the spin correlation length. However, the appropriate model to describe La$_2$O$_3$Fe$_2$Se$_2$ is not a pure $J_1-J_2$ model on a square lattice since there are two different next nearest neighbour superexchange paths: one through Se ($J_2$) and one through O ions ($J_2^{\prime}$) \cite{Neutron1}. This leads to a more complex scenario where more degenerate phases could be present and requires an accurate theoretical modelling of the system.

In conclusion, we used $^{139}$La NQR to study the local lattice modifications induced by the onset of the magnetic order in La$_2$O$_3$Fe$_2$Se$_2$ and the low-frequency dynamics developing in the paramagnetic phase. The presence of two distinct NQR peaks for $T < T_{\rm{N}}$ indicates two different structural and/or charge configurations arising in the magnetic phase. The dependence of the quadrupole coupling on temperature points at a main role of the magnetic order parameter in driving the structural distortion. The study of the spin-lattice and transverse relaxation rates shows that very low-frequency dynamics emerge in the normal phase, at frequencies in the MHz range, possibly driven by the nematic fluctuations. Future studies are needed to clarify the local differences between the two La sites.
\begin{acknowledgments}
The research in Pavia was supported by MIUR-PRIN 2015 Project No. 2015C5SEJJ. R.~H. gratefully acknowledges the financial support of the MIUR-PRIN 2015 and the Department of Physics of the University of Pavia. S.~S. acknowledges financial support from GRK-1621 graduate academy of the DFG (Project No. 129760637). S.~A. acknowledges financial support from DFG Grant No. AS 523/4-1. B.~B. acknowledges financial support from the projects of the Collaborative Research Center SFB 1143 at the TU Dresden (project-id 247310070) and Würzburg-Dresden Cluster of Excellence on Complexity and Topology in Quantum Matter–ct.qmat (EXC 2147, project-id 390858490).
\end{acknowledgments}

\providecommand{\noopsort}[1]{}\providecommand{\singleletter}[1]{#1}%

\end{document}